\documentclass[prl,aps,twocolumn,showpacs]{revtex4}
\usepackage{epsfig}
\newcommand{\ket}[1]{|#1\rangle}
\newcommand{\bra}[1]{\langle #1|}

\begin{document}
\title{The quantitative condition is necessary in guaranteeing the validity of the adiabatic approximation}
\author{D. M. Tong\footnote{Electronic address: tdm@sdu.edu.cn} }
 \affiliation{Department of Physics, Shandong University, Jinan 250100, China}
\date{\today}
\begin{abstract}
The quantitative condition has been widely used in the practical
applications of the adiabatic theorem. However, it had never been
proved to be sufficient or necessary before. It was only recently
found that the quantitative condition is insufficient, but whether
it is necessary remains unresolved. In this letter, we prove that
the quantitative condition is necessary in guaranteeing the validity
of the adiabatic approximation.
\end{abstract}
 \maketitle
\date{\today}
The adiabatic theorem reads that if a quantum system with a
time-dependent nondegenerate Hamiltonian $H(t)$ is initially in the
$n$-th instantaneous eigenstate of $H(0)$, and if $H(t)$ evolves
slowly enough, then the state of the system  at time $t$ will remain
in the $n$-th instantaneous eigenstate of $H(t)$ up to a
multiplicative phase factor. The theorem was first introduced eighty
years ago \cite{Born}, has been one of the most important theories
in quantum mechanics \cite{Schwinger,Schiff,Bohm,Kato,Messiah} and
has underpinned some of the most important developments in physical
chemistry \cite{Landau,Zener}, quantum field theory \cite{Gell},
geometric phase \cite{Berry}, and quantum computing \cite{Farhi}.
The practical applications of the theorem rely on the criterion of
the ``slowness" required by the theorem, which is usually encoded by
the quantitative condition,
\begin{eqnarray}
\left|\frac{\langle{E_n(t)}\ket{\dot E_m(t)}}{E_n(t)-E_m(t)}\right|
\ll 1,~m\neq n,~ t\in[0,\tau] \label{0}
\end{eqnarray}
where $E_m(t)$ and $\ket{E_m(t)}$ are the eigenvalues and
eigenstates of $H(t)$, and $\tau$ is the total evolution time.
Although the sufficiency as well as necessity of the condition had
been never proved before, it had been widely used as a criterion of
the adiabatic approximation. It was only recently found that the
quantitative condition is insufficient in guaranteeing the validity
of the adiabatic approximation. Marzlin and Sanders
\cite{Marzlin2004} illustrated that perfunctory application of the
adiabatic theorem may lead to an inconsistency. Tong {\sl et al}
\cite{Tong2005} pointed out that the inconsistency is a reflection
of the insufficiency of the adiabatic condition and they further
showed that the condition cannot guarantee the validity of the
adiabatic approximation. Indeed, for a given quantum system defined
by Hamiltonian $H_a(t)$ with evolution operator
$U_a(t)=\text{T}\exp(-i\int_0^tH_a(t')dt')$, one can always
construct another quantum system defined by Hamiltonian
$H_b(t)=i\dot U_a^{\dag}(t)U_a(t)$. The two systems fulfill the same
adiabatic condition, but the adiabatic approximation must be invalid
for at least one of them, which indicates that the adiabatic
condition is insufficient. These recent findings have stimulated a
great number of reexaminations on the adiabatic approximation. Some
papers contributed to the investigation of the reasons behind the
insufficiency
\cite{Vertesi2006,Duki2006,Ma2006,Ye2007,MacKenzie2007,Zhao2008,Du2008,Amin2009},
while others contributed to the development of alternative
conditions \cite{MacKenzie2006,Tong2007,Wei2007,
Jansen2007,Fujikawa2008,Wu2008,Maamache2008,Rigolin2008,Yukalov2009,Comparat2009,Huang2009,Lidar2009}
or to the examination of the validity of the quantitative condition
in concrete quantum systems
\cite{Larson2006,Liu2007,Bliokh2008,Tongpl2008,Ohara2008,Barthel
2008,Gu2009}. However, so far, whether the quantitative condition is
necessary remains unresolved. It is worth noting that some authors
have claimed that the condition was unnecessary for the adiabatic
approximation \cite{Du2008}, and it was restated in Refs.
\cite{Amin2009,Comparat2009,Lidar2009} but without a convincing
argument. Is the condition really unnecessary? It is of great
importance to put forward an exact proof. In this letter, we address
this issue. We will show that the quantitative condition defined by
Eq. (\ref{0}) is necessary in guaranteeing the validity of the
adiabatic approximation. Besides, we reexamine the spin-half model,
from which the nonnecessity was claimed, to remove the
misunderstanding on the condition.

Let us consider an $N$-dimensional quantum system with the
Hamiltonian $H(t)$. The instantaneous nondegenerate eigenvalues and
orthonormal eigenstates of $H(t)$, denoted as $E_m(t)$ and
$\ket{E_m(t)}$ respectively, are defined by
\begin{eqnarray}
H(t)\ket{E_m(t)}=E_m(t)\ket{E_m(t)},~m=1,\ldots,N. \label{1}
\end{eqnarray}
If we assume that the system is initially in the $n-$th eigenstate
$\ket{\psi(0)}=\ket{E_n(0)}$, then the state at time $t$,
$\ket{\psi(t)}$, is dictated by the Schr\"odinger equation
\begin{eqnarray}
i\frac{d}{dt}\ket{\psi(t)}=H(t)\ket{\psi(t)}. \label{2}
\end{eqnarray}
In the basis $\{\ket{E_m(t)}\}$, $\ket{\psi(t)}$ can be expanded as
\begin{eqnarray}
\ket{\psi(t)}=\sum_m c_m(t)\ket{E_m(t)}, \label{3}
\end{eqnarray}
where $c_m(t)=\bra{E_m(t)}\psi(t)\rangle$ are the time-dependent
coefficients.

We use $\ket{\psi^{adi}(t)}$ to denote the following expression,
\begin{eqnarray}
\ket{\psi^{adi}(t)}=e^{i\alpha(t)}\ket{E_n(t)},\label{padi}
\end{eqnarray}
where $\alpha(t)$ is usually written as
$\alpha(t)=-\int_0^tE_n(t')dt'+i\int_0^t\langle{E_n(t')}\ket{\dot
E_n(t')}dt'.$ In general, $\ket{\psi^{adi}(t)}$ does not fulfill the
Schr\"odinger equation, {\sl i.e.}
$i\frac{d}{dt}\ket{\psi^{adi}(t)}\neq H(t)\ket{\psi^{adi}(t)}$, and
hence it is not a solution of the Schr\"odinger equation. However,
for some quantum systems with Hamiltonians evolving slowly,
$\ket{\psi^{adi}(t)}$ may approximately fulfill the Schr\"odinger
equation, {\sl i.e.}
\begin{eqnarray}
i\frac{d}{dt}\ket{\psi^{adi}(t)}\approx
H(t)\ket{\psi^{adi}(t)}.\label{s2}
\end{eqnarray}
In this case, $\ket{\psi^{adi}(t)}$ may be taken as a good
approximation of the exact solution $\ket{\psi(t)}$, {\sl i.e.}
\begin{eqnarray}
\ket{\psi(t)}\approx\ket{\psi^{adi}(t)},\label{s3}
\end{eqnarray}
and it is said that the quantum system is in the adiabatic
evolution. This is the essential idea of the adiabatic
approximation. Note that Eq. (\ref{s2}) is necessary in ensuring
that $\ket{\psi^{adi}(t)}$ is a good approximation of the exact
solution. From Eqs. (\ref{2}), (\ref{s2}) and (\ref{s3}), we have
\begin{eqnarray}
i\frac{d}{dt}\ket{\psi(t)}&=&H(t)\ket{\psi(t)}\nonumber\\
&\approx& H(t)\ket{\psi^{adi}(t)}\nonumber\\
&\approx& i\frac{d}{dt}\ket{\psi^{adi}(t)},\label{s4}
\end{eqnarray}
which gives
\begin{eqnarray}
\ket{\dot \psi(t)}\approx \ket{\dot \psi^{adi}(t)}.\label{s5}
\end{eqnarray}
We stress that one should not take Eq. (\ref{s5}) as a trivial
result of differentiating the two sides of Eq. (\ref{s3}). Equation
(\ref{s5}) is derived from the fact that the wave function
describing the evolution of the quantum system must fulfill the
Schr\"odinger equation. In passing, we would like to mention that
Eq. (\ref{s5}) has been used in the literature by other authors, for
instance M. Berry \cite{Berry} has used it to deduce the famous
Berry phase, but here it is the first time to give a detail
discussion on its source. Besides, the validity of the adiabatic
approximation implies
\begin{eqnarray}
|c_m(t)|=\left|\bra{E_m(t)}\psi(t)\rangle\right|\ll 1, ~~m\neq n.
\label{4}
\end{eqnarray}

We now show that the condition (\ref{0}) can be deduced from Eqs.
(\ref{s3}), (\ref{s5}) and (\ref{4}). To this end, let us calculate
the coefficients $c_m(t)=\bra{E_m(t)}\psi(t)\rangle $, $m\neq n$.
Since $H(t)$ is a Hermitian operator, by using Eq. (\ref{1}), we
have $
\bra{E_m}\left(H(t)-E_n\right)\ket{\psi}=(E_m-E_n)\bra{E_m(t)}\psi(t)\rangle$.
The coefficients $c_m(t)$ can be then written as
\begin{eqnarray}
c_m(t)&=&\bra{E_m}\psi\rangle\nonumber\\
&=&\frac{1}{E_m-E_n}\bra{E_m}\left(H(t)-E_n\right)\ket{\psi},
\label{5}
\end{eqnarray}
where for abbreviation we set $E_m\equiv E_m(t)$,  $\ket{E_m}\equiv
\ket{E_m(t)}$, and $\ket{\psi}\equiv \ket{\psi(t)}$. The
Schr\"odinger equation (\ref{2}) indicates $H(t)\ket{\psi(t)}=
i\ket{\dot \psi(t)}$. Eq. (\ref{5}) can then be written as
\begin{eqnarray}
c_m(t)=\frac{1}{E_m-E_n}\bra{E_m}\left(i\ket{\dot
\psi}-E_n\ket{\psi}\right). \label{52}
\end{eqnarray}
Substituting Eqs. (\ref{s3}) and (\ref{s5}) into (\ref{52}), and
further using Eq. (\ref{padi}) and the relation
$\bra{E_m}E_n\rangle=\delta_{mn}$, we have
\begin{eqnarray}
c_m(t)&\approx& \frac{1}{E_m-E_n}\bra{E_m}\left(i\ket{\dot
\psi^{adi}}-E_n\ket{\psi^{adi}}\right)\nonumber\\
&=&\frac{e^{i\alpha}}{E_m-E_n}\bra{E_m}\left(i\ket{\dot E_n}-\dot
\alpha \ket{E_n}-E_n\ket{E_n}\right)
\nonumber\\
&=&ie^{i\alpha}\frac{\bra{E_m}\dot E_n\rangle}{E_m-E_n}. \label{6}
\end{eqnarray}
The above calculation shows that if the adiabatic approximation is
valid for the system, $c_m(t)$ must be approximately equal to
$\frac{\bra{E_m}\dot E_n\rangle}{E_m-E_n}$ up to a phase factor. In
the use of Eq. (\ref{4}), we finally obtain
$\left|\frac{\bra{E_m}\dot E_n\rangle}{E_m-E_n}\right|\ll 1$. It is
exactly the quantitative condition defined by Eq. (\ref{0}). So far,
we have completed the proof that the quantitative condition is
necessary in guaranteeing the validity of the adiabatic
approximation.

Further, we reexamine the model, a spin-half particle in a rotating
magnetic field, from which some authors claimed that the
quantitative condition was unnecessary. We will substantiate that
the quantitative condition is indeed necessary in guaranteeing the
validity of the adiabatic approximation. The Hamiltonian of the
model can be written as
\begin{eqnarray}
H(t)=\frac{\omega_0}{2}(\sigma_x\sin\theta\cos\omega
t+\sigma_y\sin\theta\sin\omega t+\sigma_z\cos\theta),
\end{eqnarray}
where $\omega_0$ is a time-independent parameter defined by the
magnetic moment of the spin and the intensity of external magnetic
field, $\omega$ is the rotating frequency of the magnetic field and
$\sigma_i,~i=x,y,z,$ are Pauli matrices. Without loss of generality,
we suppose $\omega_0>0$, $\omega>0$, and $\sin\theta\neq 0$. The two
instantaneous eigenvalues of $H(t)$ are $ E_1=-\frac{\omega_0}{2},~
E_2=\frac{\omega_0}{2}$,  and the instantaneous eigenstates are
\begin{eqnarray}
\ket{E_1(t)}=\left(\begin{array}{c} e^{-i\omega
t/2}\sin\frac{\theta}{2}\\-e^{i\omega t/2}\cos\frac{\theta}{2}
\end{array}\right),~~
\ket{E_2(t)}=\left(\begin{array}{c} e^{-i\omega
t/2}\cos\frac{\theta}{2}\\e^{i\omega t/2}\sin\frac{\theta}{2}
\end{array}\right),
\label{ketE2}
\end{eqnarray}
respectively. The Schr\"odinger equation for the model reads
\begin{eqnarray}
i\frac{d}{dt}\ket{\psi(t)}=\frac{\omega_0}{2}\left(\begin{array}{cc}
\cos\theta&\sin\theta e^{-i\omega t}\\
\sin\theta e^{i\omega t}&-\cos\theta\end{array}\right)\ket{\psi(t)}.
\label{schrodinger2}
\end{eqnarray}
Suppose that the system is initially in the first eigenstate,
$\ket{\psi(0)}=\ket{E_1(0)}$. In the basis $\ket{E_1(t)}$ and
$\ket{E_2(t)}$, $\ket{\psi(t)}$ can be expanded as
\begin{eqnarray}
\ket{\psi(t)}=a(t)\ket{E_1(t)}+b(t)\ket{E_2(t)}, \label{solution2}
\end{eqnarray}
where $a(t)$, $b(t)$ are two time-dependent coefficients to be
determined. Substituting Eq. (\ref{solution2}) into
(\ref{schrodinger2}), we may obtain the differential equations
fulfilled by $a(t)$ and $b(t)$, from which we have
\begin{eqnarray}
a(t)&=&\left(\cos\frac{\overline{\omega}t}{2}+i\frac{\omega_0-\omega\cos\theta}{\overline{\omega}}
\sin\frac{\overline{\omega}t}{2}\right),\nonumber
\\
b(t)&=&i\frac{\omega\sin\theta}{\overline{\omega}}\sin\frac{\overline{\omega}t}{2},
\label{atbt}
\end{eqnarray}
with
$\overline{\omega}=\sqrt{\omega_0^2+\omega^2-2\omega_0\omega\cos\theta}$.

For this model, the quantitative condition is  $ \omega_0\gg
\omega\sin\theta$, and
$\ket{\psi^{adi}(t)}=e^{\frac{i}{2}\omega_0t}\ket{E_1(t)}$. If the
adiabatic approximation is valid, there must be
\begin{eqnarray}
\left|b(t)\right|\sim\frac{\omega\sin\theta}{\sqrt{\omega_0^2+\omega^2-2\omega_0\omega\cos\theta}}\ll
1.\label{111}
\end{eqnarray}
For convenience's sake, we denote the term on the left-hand side of
Eq. (\ref{111}) by $f(\frac{\omega_0}{\omega})$, and take it as a
function of $\frac{\omega_0}{\omega}$, i.e. $
f(\frac{\omega_0}{\omega})=\frac{\sin\theta}{\sqrt{(\frac{\omega_0}{\omega})^2-2
(\frac{\omega_0}{\omega})\cos\theta+1}}$.
We now analyze the values of $f(\frac{\omega_0}{\omega})$.
Noting that the sign of $\cos\theta$ changes from positive to
negative at $\theta=\pi/2$, we pursue the discussions respectively
for $0<\theta\leq\pi/2$ and for $\pi/2<\theta<\pi$. In the first
case where $\theta\in(0,\frac{\pi}{2}]$,
$f(\frac{\omega_0}{\omega})$ is a monotonic increasing function for
$\frac{\omega_0}{\omega}<\cos\theta$ and a monotonic decreasing
function for $\frac{\omega_0}{\omega}> \cos\theta$. It has a maximum
at $\frac{\omega_0}{\omega}=\cos\theta$.
$f(\frac{\omega_0}{\omega})$ is always larger than $\sin\theta$ in
the interval $0<\frac{\omega_0}{\omega}\leq\cos\theta$ and larger
than $\sin\frac{\theta}{2}$ in the interval
$\cos\theta<\frac{\omega_0}{\omega}\leq 1$. The solid line in Fig.
$1$ is a sketch of $f(\frac{\omega_0}{\omega})$ for
$\theta\in(0,\frac{\pi}{2}]$.
\begin{figure}[htbp]
\begin{center}
\includegraphics[width=8cm, height=4cm]{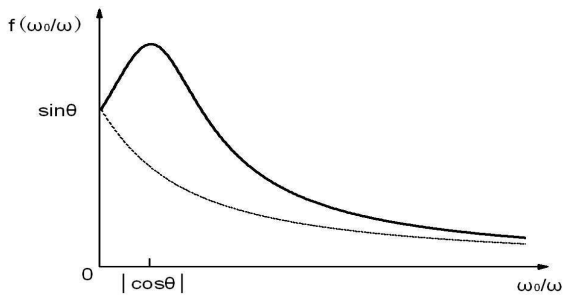}
\end{center}
\caption{A sketch of $f\left(\frac{\omega_0}{\omega}\right)$. The
solid line is for $\theta\in(0,\frac{\pi}{2}]$ and the dashed line
is for $\theta\in(\frac{\pi}{2},\pi)$. \label{casel}}
\end{figure}
In the second case where $\theta\in(\frac{\pi}{2},\pi)$,
$f(\frac{\omega_0}{\omega})$ is a monotonic decreasing function of
$\frac{\omega_0}{\omega}$ in its domain. It is always larger than
$\sin\frac{\theta}{2}$ in the interval
$0<\frac{\omega_0}{\omega}\leq 1$. The dashed line in Fig. $1$ is a
sketch of $f(\frac{\omega_0}{\omega})$ for
$\theta\in(\frac{\pi}{2},\pi)$. These calculations show that for a
nonzero $\sin\theta$, the adiabatic approximation is valid only if
$\omega_0\gg\omega$, which necessarily implies the quantitative
condition $\omega_0\gg \omega\sin\theta$. If $\omega_0\gg\omega$ is
not fulfilled, for instance $\omega_0\ll\omega$ or
$\omega_0\sim\omega$ , the absolute value of $b(t)$ in Eq.
(\ref{solution2}) is in the order of $ \sin\theta$, and therefore
the adiabatic approximation is invalid.

After having demonstrated that $\omega_0\gg \omega\sin\theta$ is a
necessary condition for the adiabatic evolution of the spin-half
system, we now explain
what is wrong in the claim that the quantitative condition was
unnecessary. It was argued that if $\sin\theta$ is small enough, the
fidelity between $\ket{\psi(t)}$ and $\ket{\psi^{adi}(t)}$ will then
be close to $1$ and the adiabatic approximation would be valid even
if $\omega_0\ll\omega$. Certainly, it is true that the fidelity may
be close to $1$ if $\sin\theta$ is small enough, but this does not
imply that the adiabatic approximation is valid for
$\omega_0\ll\omega$. In fact, $\ket{\psi(t)}$ cannot be expressed as
$a(t)\ket{E_1(t)}$ if only $\sin\theta$ is small but not
$\omega_0\gg\omega$. To clarify this point, let us rewrite Eq.
(\ref{solution2}) as $ \ket{\psi(t)}=\left(\begin{array}{c}
A_1+B_1\\A_2+B_2
\end{array}\right)$, where $A_i$ and $B_i$ are determined by
$\left(\begin{array}{c} A_1\\A_2
\end{array}\right)\equiv a(t)\ket{E_1(t)}, ~\left(\begin{array}{c} B_1\\B_2
\end{array}\right)\equiv b(t)\ket{E_2(t)}$.
By using Eqs (\ref{ketE2}),(\ref{solution2}) and (\ref{atbt}), the
explicit expressions of $A_i$ and $B_i$ can be obtained. One may
find that $B_2$ relative to $A_2$ is much smaller, and it is valid
to have $A_2+B_2\approx A_2$. Yet, $B_1$ is of the same order as
$A_1$, and it is invalid to take $A_1+B_1\approx A_1$. Therefore,
one cannot take $a(t)\ket{E_1(t)}$ as an approximation of
$\ket{\psi(t)}$. Further more, we can also find the distinct
difference between $\ket{\psi(t)}$ and $\ket{\psi^{adi}(t)}$ by
comparing the Bloch vectors of them. The exact solution
(\ref{solution2}) can be explicitly written as
\begin{eqnarray}
\ket{\psi(t)}=\left(\begin{array}{c} e^{-i\omega
t/2}\sin\frac{\theta}{2}\left(\cos\frac{\overline{\omega}t}{2}+
i\frac{\omega_0+\omega}{\overline\omega}\sin\frac{\overline{\omega}t}{2}\right)\\
-e^{i\omega
t/2}\cos\frac{\theta}{2}\left(\cos\frac{\overline{\omega}t}{2}+i\frac{\omega_0-\omega}{\overline\omega}\sin\frac{\overline{\omega}t}{2}\right)
\end{array}\right).
\label{bloch1}
\end{eqnarray}
If  $\omega_0\ll\omega$, we have
$\frac{\omega_0\pm\omega}{\overline\omega}\approx \pm 1$ and
$\overline\omega\approx \omega-\omega_0\cos\theta+\delta$, where
$\delta=\delta(\frac{\omega_0}{\omega})$ is of the order
$\frac{\omega_0}{\omega}$. Equation (\ref{bloch1}) then becomes
\begin{eqnarray}
\ket{\psi(t)}\approx \left(\begin{array}{c}
e^{-i(\omega_0\cos\theta-\delta)
t/2}\sin\frac{\theta}{2}\\
-e^{i(\omega_0\cos\theta-\delta)t/2}\cos\frac{\theta}{2}
\end{array}\right).
\label{bloch2}
\end{eqnarray}
Clearly, the Bloch vector of $\ket{\psi^{adi}(t)}$  is rotating as
fast as the magnetic field. However, from Eq. (\ref{bloch2}), we
find that for the exact solution $\ket{\psi(t)}$, the rotating rate
of its Bloch vector is about $\omega_0$, which is far from the
rotating rate of the magnetic field. Therefore, if
$\omega_0\ll\omega$, the system is never in the adiabatic evolution,
no matter how small $\sin\theta$ is. For instance, if we take
$\theta=0.06$ and $\omega=10\omega_0$, as in Ref. \cite{Du2008}, the
rotating rate of the state $\ket{\psi(t)}$ is 10 times as much as
that of $\ket{\psi^{adi}(t)}$ although the fidelity between the two
states is close to $1$.

In summary, we have proved that the quantitative condition is
necessary in guaranteeing the validity of the adiabatic
approximation. One can then conclude that the quantitative condition
is a necessary but insufficient one. Fulfilling only the
quantitative condition may not guarantee the validity of the
adiabatic approximation, but violating the condition must lead to
the invalidity of the approximation. Since the quantitative
condition plays an important role in the practical applications of
the adiabatic theorem and it had been found to be insufficient, the
confirmation of its necessity is of great importance. Besides, the
findings in the letter have removed all the previous doubts or
misunderstandings on the quantitative condition. In passing, we
would like to point out that the quantitative condition may be a
necessary and sufficient criterion of the adiabatic approximation
for a large number of interesting quantum systems, although it is
difficult to pick out these systems. This may be the underlying
reason that the quantitative condition is still a powerful tool
widely used by researchers despite the finding of its insufficiency.

\vskip 0.3 cm D. M. Tong thanks G. L. Long and J. F. Du for useful
discussions. This work was supported by NSF China with No.10875072
and the National Basic Research Program of China (Grant No.
2009CB929400).


\begin{thebibliography}{99}
\bibitem{Born}M. Born and V. Fock, Z. Phys. {\bf 51}, 165(1928).
\bibitem{Schwinger}J. Schwinger, Phys. Rev. {\bf 51}, 648(1937).
\bibitem{Schiff}L. I. Schiff, Quantum Mechanics (McGRAW-Hill Book Co., Inc., New York, 1949).
\bibitem{Bohm}D. Bohm, Quantum Theory (Prentic-Hall, Inc., New York, 1951).
\bibitem{Kato}T. Kato, J. Phys. Soc. Jap. {\bf 5}, 435 (1950).
\bibitem{Messiah}A. Messiah, Quantum Mechanics (North-Holland Pub. Co., Amsterdam,1962).
\bibitem{Landau}L. D. Landau, Phys. Z. Sowjetunion {\bf 2}, 46
(1932).
\bibitem{Zener} C. Zener, Proc. R. Soc. London A {\bf 137}, 696
(1932).
\bibitem{Gell} M. Gell-Mann and F. Low, Phys. Rev. {\bf 84}, 350
(1951).
\bibitem{Berry} M.V. Berry, Proc. R. Soc. London Ser. A {\bf 392}, 45
(1984).
\bibitem{Farhi} E. Farhi {\it et al}, Science  {\bf 292}, 472(2001).
\bibitem{Marzlin2004}K. P. Marzlin and B. C. Sanders, Phys. Rev. Lett. {\bf 93}, 160408 (2004).
\bibitem{Tong2005}D. M. Tong, K. Singh, L. C. Kwek, and C. H. Oh, Phys. Rev. Lett. {\bf 95}, 110407 (2005).


\bibitem{Vertesi2006} T. Vertesi, R. Englman,  Phys. Lett. A {\bf 353}, 11 (2006).
\bibitem{Duki2006} S. Duki, H. Mathur, O. Narayan, Phys. Rev. Lett. {\bf 97}, 128901 (2006).
\bibitem{Ma2006} J. Ma, Y. P. Zhang, E. G. Wang, B. Wu, Phys. Rev. Lett. {\bf 97}, 128902 (2006).
\bibitem{Ye2007} M. Y. Ye, X. F. Zhou, Y. S. Zhang {\it et al.},  Phys. Lett. A {\bf 368}, 18 (2007).
\bibitem{MacKenzie2007} R. MacKenzie, A. Morin-Duchesne, H. Paquette,J. Pinel/, Phys. Rev. A {\bf 76}, 044102 (2007).
\bibitem{Zhao2008} Y. Zhao, Phys. Rev. A {\bf 77}, 032109 (2008).
\bibitem{Du2008} J. F. Du, L. Z. Hu, Y. Wang {\it et al.}, Phys. Rev. Lett. {\bf 101}, 060403 (2008).
\bibitem{Amin2009} M. H. S. Amin, Phys. Rev. Lett.  {\bf 102}, 220401 (2009).

\bibitem{MacKenzie2006} R. MacKenzie, E. Marcotte, H. Paquette, Phys. Rev. A {\bf 73}, 042104 (2006).
\bibitem{Tong2007}D. M. Tong, K. Singh, L. C. Kwek, C. H. Oh, Phys. Rev. Lett. {\bf 98}, 150402 (2007).
\bibitem{Wei2007} Z. H. Wei, M.S. Ying, Phys. Rev. A {\bf 76}, 024304 (2007).
\bibitem{Jansen2007}S. Jansen, M. B.Ruskai, R. Seiler, J. Math. Phys. {\bf 48}, 102111 (2007).
\bibitem{Fujikawa2008} K. Fujikawa, Phys. Rev. D {\bf 77}, 045006 (2008).
\bibitem{Wu2008} J. D. Wu, M .S. Zhao, J. L. Chen, Y. D. Zhang, Phys. Rev. A {\bf 77}, 062114 (2008).
\bibitem{Maamache2008} M. Maamache, Y. Saadi, Phys. Rev. Lett. {\bf 101}, 150407 (2008).
\bibitem{Rigolin2008} G. Rigolin, G. Ortiz, V. H. Ponce, Phys. Rev. A {\bf 78}, 052508 (2008).
\bibitem{Yukalov2009} V. I. Yukalov, Phys. Rev. A {\bf 79}, 052117 (2009).
\bibitem{Comparat2009} D. Comparat, Phys. Rev. A {\bf 80}, 012106 (2009).
\bibitem{Huang2009} X. L. Huang, X. X. Yi, Phys. Rev. A {\bf 80}, 032108 (2009).
\bibitem{Lidar2009} D. A. Lidar, A. T. Rezakhani, A. Hamma, J Math. Phys. {\bf 50}, 102106 (2009).

\bibitem{Larson2006} J. Larson  and S. Stenholm , Phys. Rev.  A {\bf 73}, 033805 (2006).
\bibitem{Liu2007} J. Liu and L. B. Fu, Phys. Lett. A {\bf 370}, 17 (2007).
\bibitem{Bliokh2008} K. Y. Bliokh, Phys. Lett.  A {\bf 372}, 204 (2008).
\bibitem{Tongpl2008} D. M. Tong, X. X. Yi {\it et al.}, Phys. Lett. A  {\bf 372}, 2364 (2008).
\bibitem{Ohara2008}M. J. O'Hara and D. P. O'Leary, Phys. Rev. A {\bf 77}, 042319 (2008).
\bibitem{Barthel 2008} T. Barthel, C. Kasztelan, I. P. McCulloch, U. Schollwock/, Phys. Rev. A {\bf 79}, 053627 (2009).
\bibitem{Gu2009} S. J. Gu, Phys. Rev. E {\bf 79}, 061125 (2009).


\end{thebibliography}
\end{document}